\documentclass[trackchanges,twocolumn]{aastex7}

\usepackage{subcaption}

\usepackage{amsmath}

\begin{document}

\title{JWST NIRCam Observations of the Globular Cluster Population of RXJ 2129.7+0005}




\author[orcid=0009-0001-7599-1967,sname='Keatley']{Kaitlyn E. Keatley}
\affiliation{Department of Physics \& Astronomy, McMaster University, 1280 Main St W, Hamilton, ON L8S 4L8, Canada}
\email{keatleyk@mcmaster.ca}

\author[orcid=0000-0001-8762-5772, sname='Harris']{William E. Harris}
\affiliation{Department of Physics \& Astronomy, McMaster University, 1280 Main St W, Hamilton, ON L8S 4L8, Canada}
\email{harrisw@mcmaster.ca}

\begin{abstract}

We present an analysis of the globular cluster (GC) population in the galaxy cluster RXJ 2129.7+0005 (z = 0.234) based on JWST NIRCam imaging in three filters: F115W, F150W, and F200W. We use this material to provide a detailed look at the color-magnitude distribution of the GCs and their spatial distribution around the central giant galaxy. We identified 3,160 GC candidates brighter than F150W=29.5, and assessed photometric completeness through artificial star tests. We determined that the GCs follow a radial power-law distribution with an index of $1.58 \pm 0.04$, with the redder GCs exhibiting a slightly greater central concentration.
Their spatial distribution is also highly elliptical, closely following the shape of the BCG halo light.

\end{abstract}




\section{Introduction} \label{sec:intro}

Globular clusters (GCs) present numerous avenues for exploring the evolutionary histories of galaxies and galaxy clusters. GCs distribute themselves according to the gravitational potential of their host galaxies and galaxy clusters, offering insights into the history of the system.
Given that galaxies in clusters reside within massive dark matter halos that dominate the gravitational field, GCs have great potential as a tool to probe these halos. Studies such as \cite{Blakeslee1997} and \cite{Harris2017} have demonstrated strong correlations between the mass of GC systems and the mass of the dark matter halo, reaffirming that regions with higher mass are expected to host more GCs.

Using simulations and kinematic models, there have been some promising results in probing the dark matter halo using the diffuse stellar halo (e.g. \cite{pillepich2017}) and satellite galaxies (e.g. \cite{Alabi2016}, \cite{Slizewski2022}). 
However, these approaches face challenges: observationally measuring diffuse stellar halos is difficult at large radii, and satellite galaxy studies require highly accurate spectroscopy.
In contrast, GCs are more numerous than satellite galaxies, and brighter and more extended than the diffuse stellar halo, making them highly promising tracers of dark matter in galaxy clusters.

Prior to JWST, the available resolution and sensitivity were insufficient to effectively study GC populations beyond the local universe. 
We have only begun to utilize the remarkable resolution and depth of JWST to study whole GC populations at intermediate redshifts. 
Notable examples include VV 191a \citep{Berkheimer2024}, Abell 2744 \citep{Harris2023, Harris2024}, and SMACS J0723.3–7327 \citep{Lee2022}.
These rich databases allow us to directly study ancient star clusters at earlier periods in their evolution, enabling new insights into both GC and galaxy cluster evolution.

RXJ 2129.7+0005 (hereafter referred to as RXJ 2129) is a virialized galaxy cluster at a redshift of 0.234 corresponding to a lookback time of 2.90 Gyr. It has a central BCG, a mini radio halo and cool core \citep{Kale2015,Giacintucci2017,Ueda2020}, and an x-ray cold front in the intracluster light (ICL) demonstrating a rich merger history \citep{jimenezteja2024}.
Featuring several strong lensing features, RXJ 2129 was a target of the Cluster Lensing And Supernova survey with Hubble (CLASH, \cite{Postman2012}), leading to many strong and weak lensing studies on its mass distribution (\cite{Desprez2018,Umetsu2018,Caminha2019,Jauzac2021,caminha2022}). The GC system has not previously been studied.

In this paper, we utilize new high resolution JWST NIRCam images of RXJ 2129 to analyze the spatial distribution of the GCs and begin a preliminary investigation into comparisons to mass models of the galaxy cluster. 
This is part of a series of studies of GCs in lensing clusters that begun with Abell 2744 \citep{Harris2023, Harris2024}. 
In Section \ref{sec:data}, we describe our photometry procedure for detecting GC candidates and determining their magnitudes. We also perform artificial star tests to test the completeness and photometric limits of our measurements. In Sections \ref{sec:radial} and \ref{sec:color}, we determine the radial distribution of GCs, and investigate any trends with GC color. Additionally, we explore the angular distribution in Section \ref{subsec:ellipticity}, determining the significant ellipticity of the BCG. In section \ref{sec:abell2744}, we compare the color-magnitude diagram and predicted metallicity to that of Abell 2744. Finally, in Section \ref{sec:conclusion}, we summarize our findings and outline the additional work that is ongoing with RXJ 2129 and distant galaxy clusters like it. 

The luminosity distance to RXJ 2129 is 1.20 Gpc, with cosmological parameters of $H_0=67.8$ km/s and $\Omega_\Lambda=0.692$ \citep{planck}. 

\section{Data} \label{sec:data}

\begin{figure}[ht!]
\centering
\includegraphics[width=0.42\textwidth]{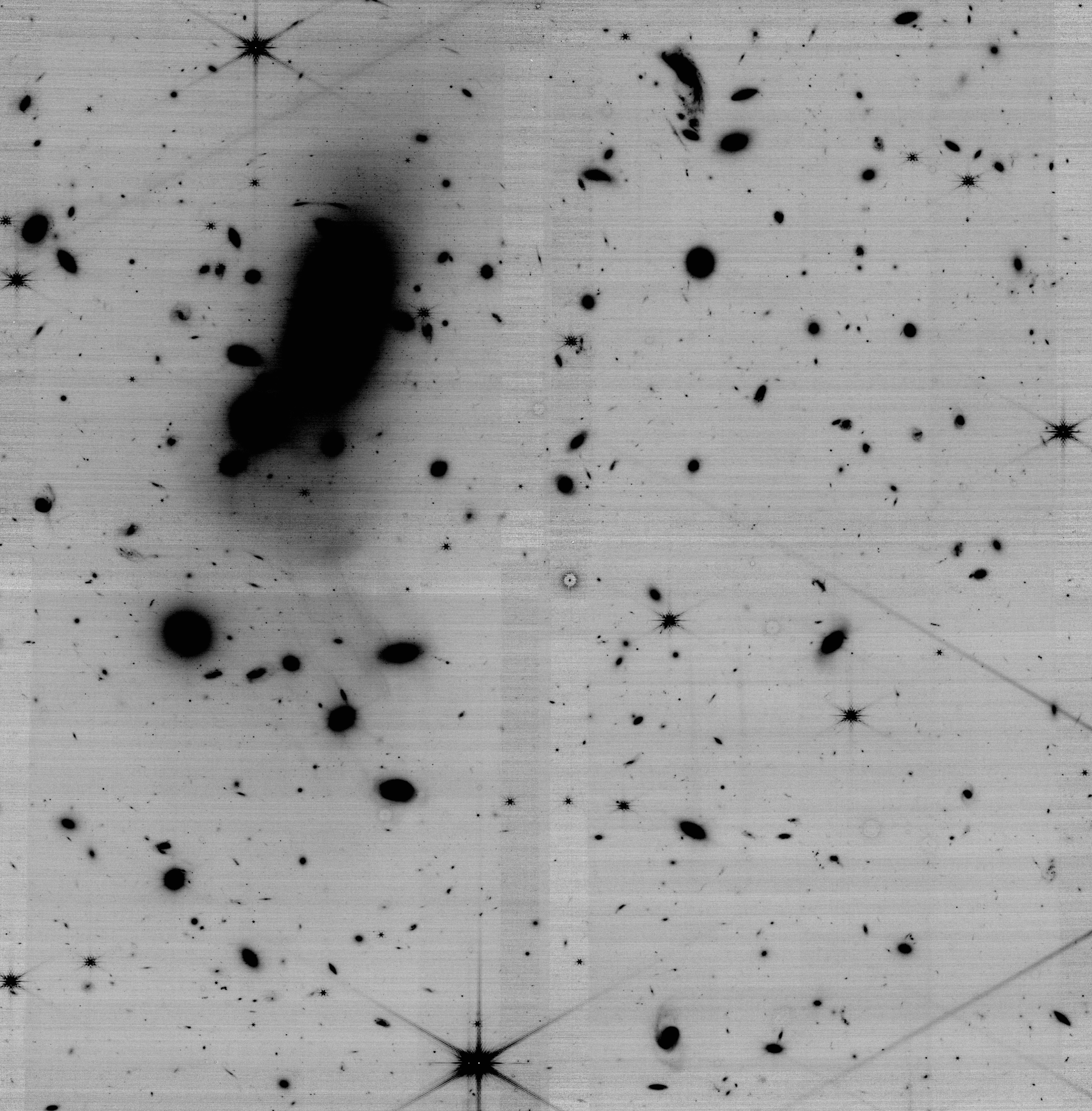}
\caption{F200W view of RXJ 2129. The BCG is the in the upper left. Smaller galaxies and foreground stars are also visible. Image is 139.6 x 138.9 arcsec, which is equivalent to 534 kpc across.  North is at top and East is to the left.
\label{fig:field}}
\end{figure}

We use images from JWST NIRCam in the F115W, F150W, and F200W filters drawn from the MAST JWST archive (DD 2767, PI P. Kelly). 
The F115W and F200W images had an exposure time of 8246 seconds, and the F150W image had an exposure time of 19927 seconds.
The camera scale is 0.0307 arcsec per pixel, with a total image size of 139.6 x 138.9 arcsec. Fig \ref{fig:field} shows a portion of the field with the BCG. 

\subsection{Photometry and Object Selection}

We conduct photometry and object selection using the \texttt{daophot} software \citep{daophot} within IRAF. First, we add the three images together to make a summed image, and then apply a median filter to eliminate large-scale gradients. With a detection threshold of $4 \sigma_s$ where $\sigma_s$ is the standard deviation of the sky noise, we use the \textit{daofind} package on this median-subtracted image to find objects.
The summation yields a combined image that has relatively lower sky noise and allows for a deeper finding list of objects.
For each of the three images, we use \textit{phot} to conduct photometry using a 3-px aperture radius for each object on the finding list. A point spread function (PSF) is constructed for each filter using 46 foreground stars and a FWHM of 2.0, 2.4, and 2.8 px for the F115W, F150W, and F200W images, respectively. 
The interactive interface of the \textit{psf} package allows us to inspect the profile of each candidate star to hand-pick unsaturated, isolated foreground stars. We then run the \textit{allstar} function to get final instrumental magnitudes on all star-like objects in our photometry list. 

\begin{figure}[ht!]
\includegraphics[width=0.5\textwidth]{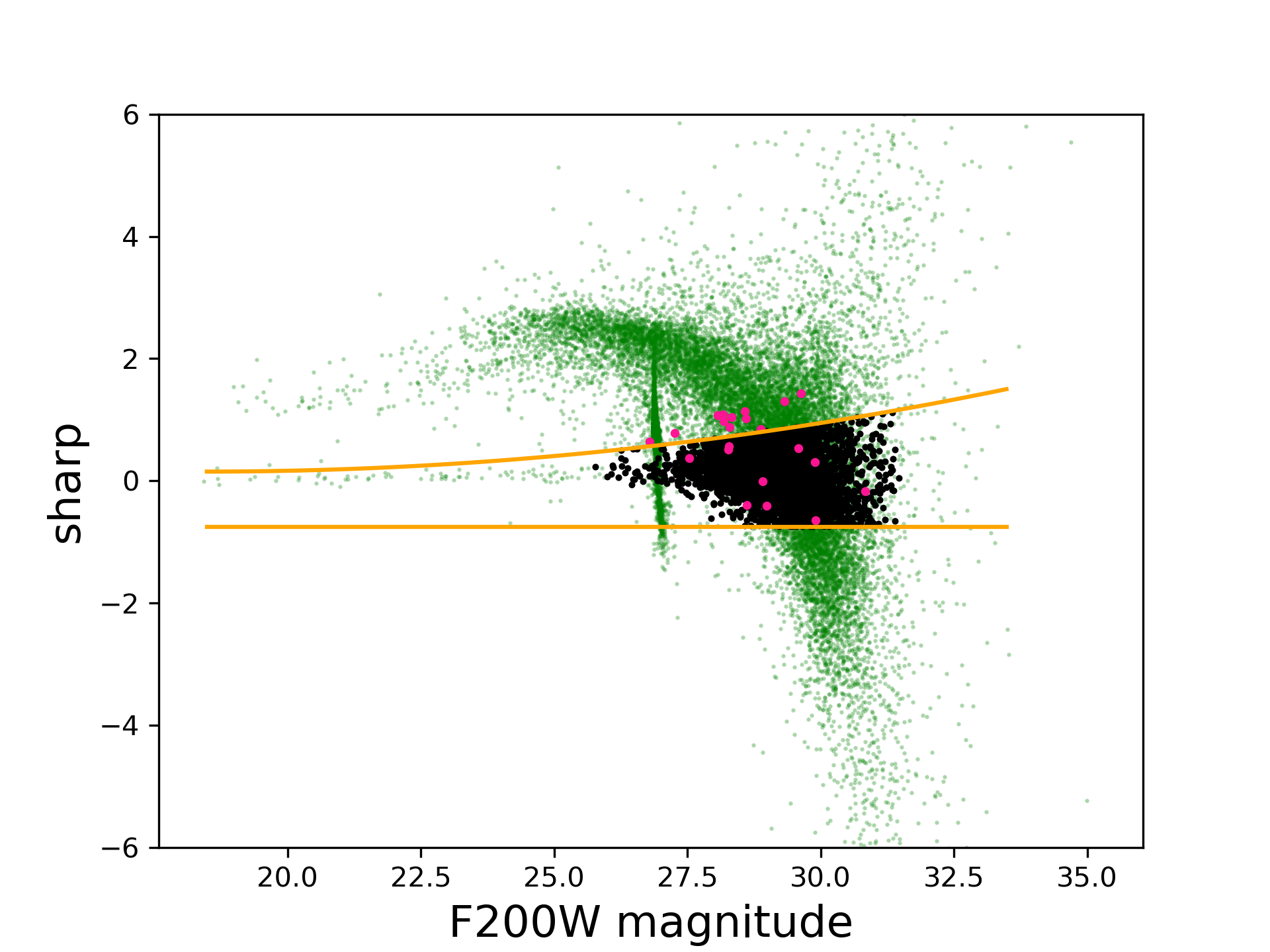}
\caption{Sharp parameter of detected objects versus magnitude in AB magnitudes. All detected objects are shown as green points. Accepted points using the sharp and chi cuts are shown in black, and selected point sources in the innermost region of the BCG are shown in pink.
\label{fig:sharp}}
\end{figure}

\textit{Allstar} additionally provides \textit{sharp} and \textit{chi} parameters, which we use to isolate point sources by plotting these quantities against the F150W and F200W magnitudes. 
We make a quadratic cut in sharp-mag space shown as orange boundaries in Fig \ref{fig:sharp}. 
We reject points with sharp (in F150W or F200W) less than -0.75, or lying above the line $s=0.006(m-8)^2+0.15$
where $s$ is the sharp parameter and m is the uncorrected magnitude. 
We also remove any object that has a chi value larger than 4 in any filter. 
Accepted point sources are shown as black points in Fig \ref{fig:sharp}. Additional objects near the center of the BCG were independently selected by close visual inspection and are shown as pink points.

\begin{figure*}[ht!]
\centering
  \begin{subfigure}{0.45\textwidth}
    \includegraphics[width=\textwidth]{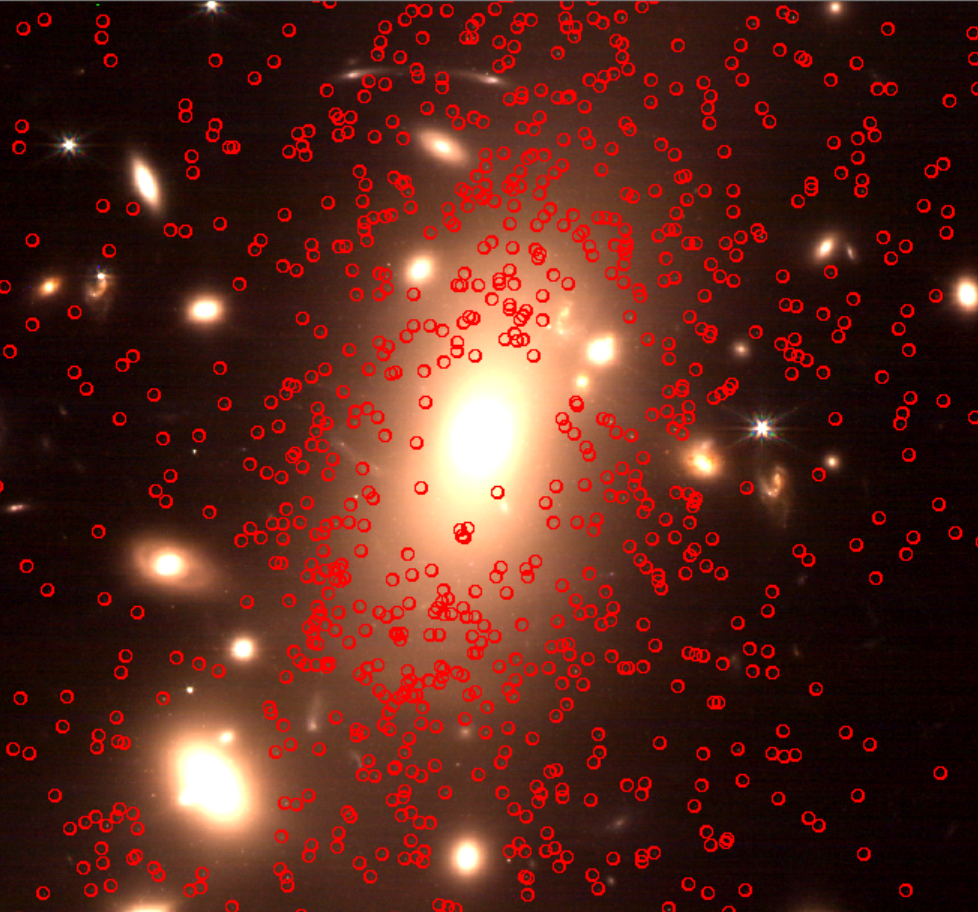}
  \end{subfigure}
  \hfill
  \begin{subfigure}{0.45\textwidth}
    \includegraphics[width=\textwidth]{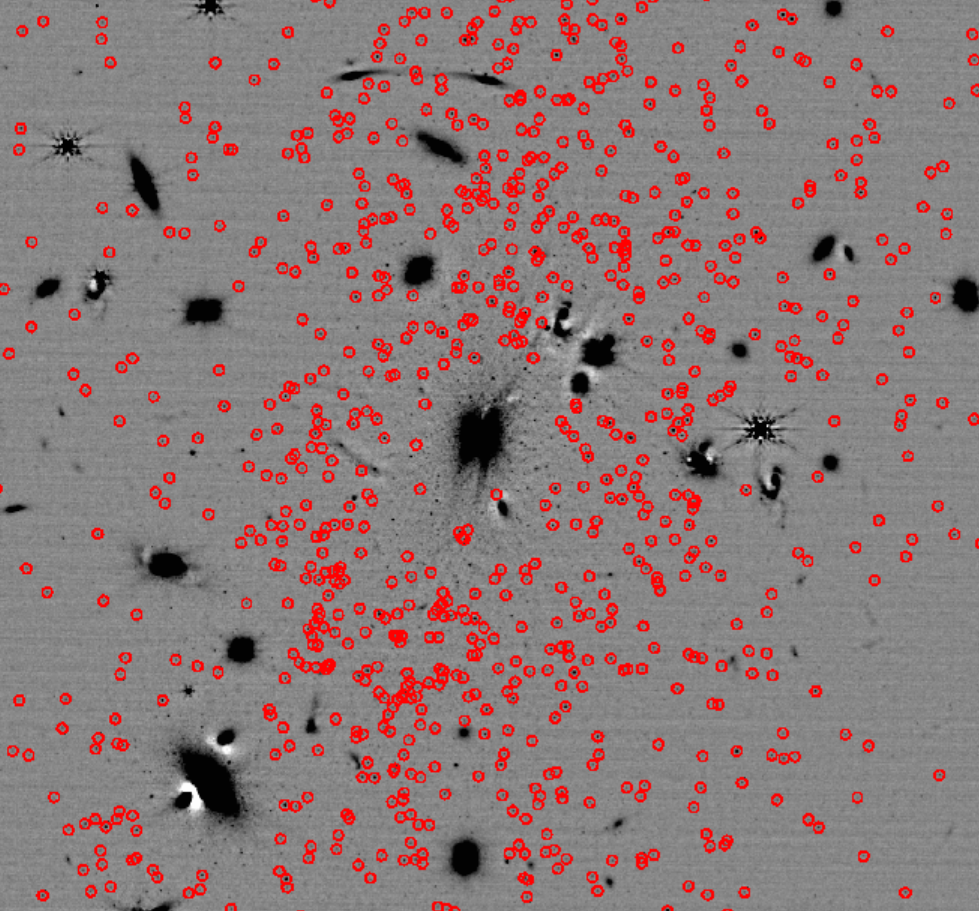}
  \end{subfigure}
\caption{Color image (left) and unsharp-masked F200W image (right) of the field near the BCG of RXJ 2129. Point sources are circled in red. Box size is 37.3 x 34.9 arcsec, equivalent to 143 x 133 kpc.
\label{fig:selected}}
\end{figure*}

In total, 4642 objects are recovered in all three filters using the combined finding list. Recovered point sources near the BCG are circled in red in Fig \ref{fig:selected}. 

To convert magnitudes from the \textit{allstar} scale to AB magnitudes, we first correct PSF \textit{allstar} magnitudes to large aperture using bright isolated stars with a 10-px aperture radius.
To integrate to `infinite' radius, we use the PSF encircled energy from the NIRCam webpage\footnote{https://jwst-docs.stsci.edu/jwst-near-infrared-camera/nircam-performance/nircam-point-spread-functions}.
Cosmological K-corrections are applied for a redshift of 0.234 using the webtool \texttt{RESCUER} \citep{kcorr}.
Finally, we adjust our magnitudes for foreground reddening (NED database).
Foreground absorption for RXJ 2129 is
0.040, 0.026, and 0.017; K-corrections are -0.137, -0.130, -0.370 for the F115W, F150W, and F200W filters, respectively. 

\subsection{Completeness fraction}

GCs that are very faint or located in regions of high sky noise will naturally be more difficult to detect than bright, isolated GCs. 
To determine the completeness of our selection, we perform artificial star tests. Using the \textit{addstar} function, we add 1060 artificial stars into each image. These stars are distributed radially to mirror the radial distribution of GCs. 
We use logistic regression fitting (Equation \ref{eqn:lr}) to determine the recovery probability as a function of magnitude and local sky noise, as described in \cite{logregression} and \cite{Harris2024}:

\begin{equation}
    p(m) = \frac{1}{(1+e^{-g(x_i)})}
    \label{eqn:lr}
\end{equation}

where $g(x_i) = \beta_0 + \beta_1 x_1 + \beta_2 x_2$. Here, $x_1$ is the magnitude and $x_2$ is the local sky noise of each object, and the $\beta_n$ values are determined by the fit using the artificial stars. The local sky noise is defined as the standard deviation of the sky pixel intensities in an annulus of 0.31 to 0.46 arcsec radii.

\begin{figure}[ht!]
\includegraphics[width=0.5\textwidth]{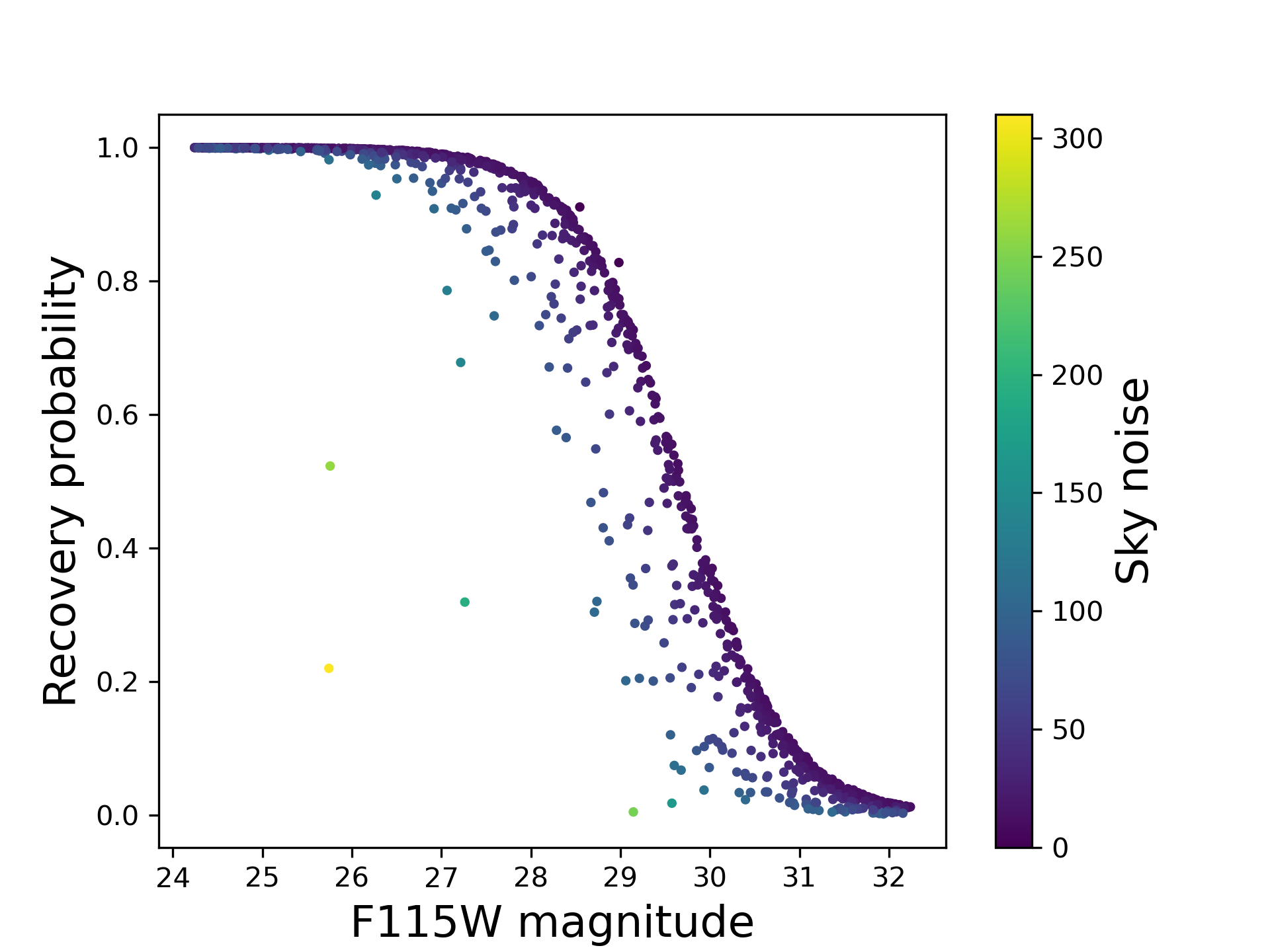}
\caption{Recovery probability in the F115W filter of artificial stars using a logistic regression fit. Points are color-coded by the local sky noise.  Points scattering below the main line are those in regions of higher sky background. 
\label{fig:recov_prob}}
\end{figure}

A maximum-likelihood solution was performed in Python using
\textit{statsmodels/scikitlearn}. 
Recovery probabilities of the 1060 artificial stars using the logistic regression fit are shown in Fig \ref{fig:recov_prob}. We obtain coefficient values of $\beta_0=52 \pm 3$, $\beta_1= -1.7\pm 0.1$, and $\beta_2=-0.027\pm 0.004$ using corrected input magnitudes, which we use to determine the recovery probability for each of our GC candidates. We define our photometric completeness limit as the magnitude at which $p=0.5$.

\begin{figure*}[ht!]
\centering
  \begin{subfigure}{0.95\textwidth}
    \includegraphics[width=\textwidth]{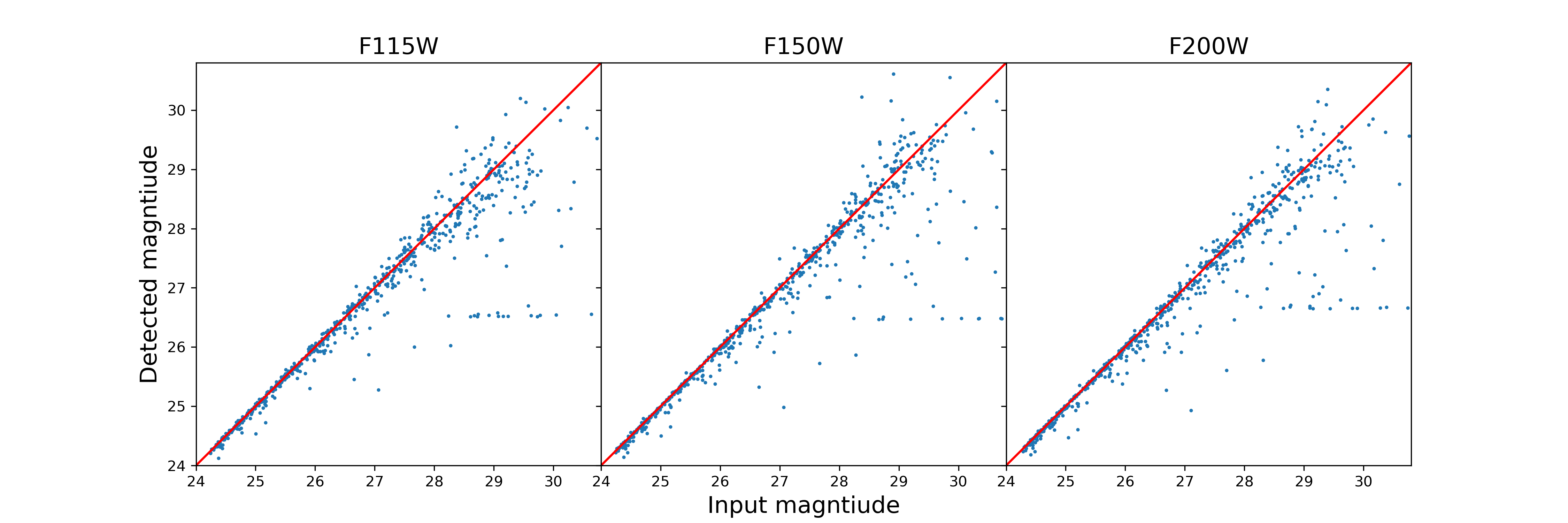}
  \end{subfigure}
  \hfill
\caption{Input versus detected magnitude for 1060 artificial stars in the F115W, F150W, and F200W images. The red line shows a 1-to-1 relationship. 
\label{fig:input_det}}
\end{figure*}

As an additional check, we also compare the input and detected magnitudes of the artificial stars to check for any systematic biases. As shown in Fig \ref{fig:input_det}, there is a slight tendency for brighter magnitudes to be recovered for the fainter objects, but is negligible at magnitudes brighter than the completeness limit.

\begin{figure}[ht!]
\includegraphics[width=0.49\textwidth]{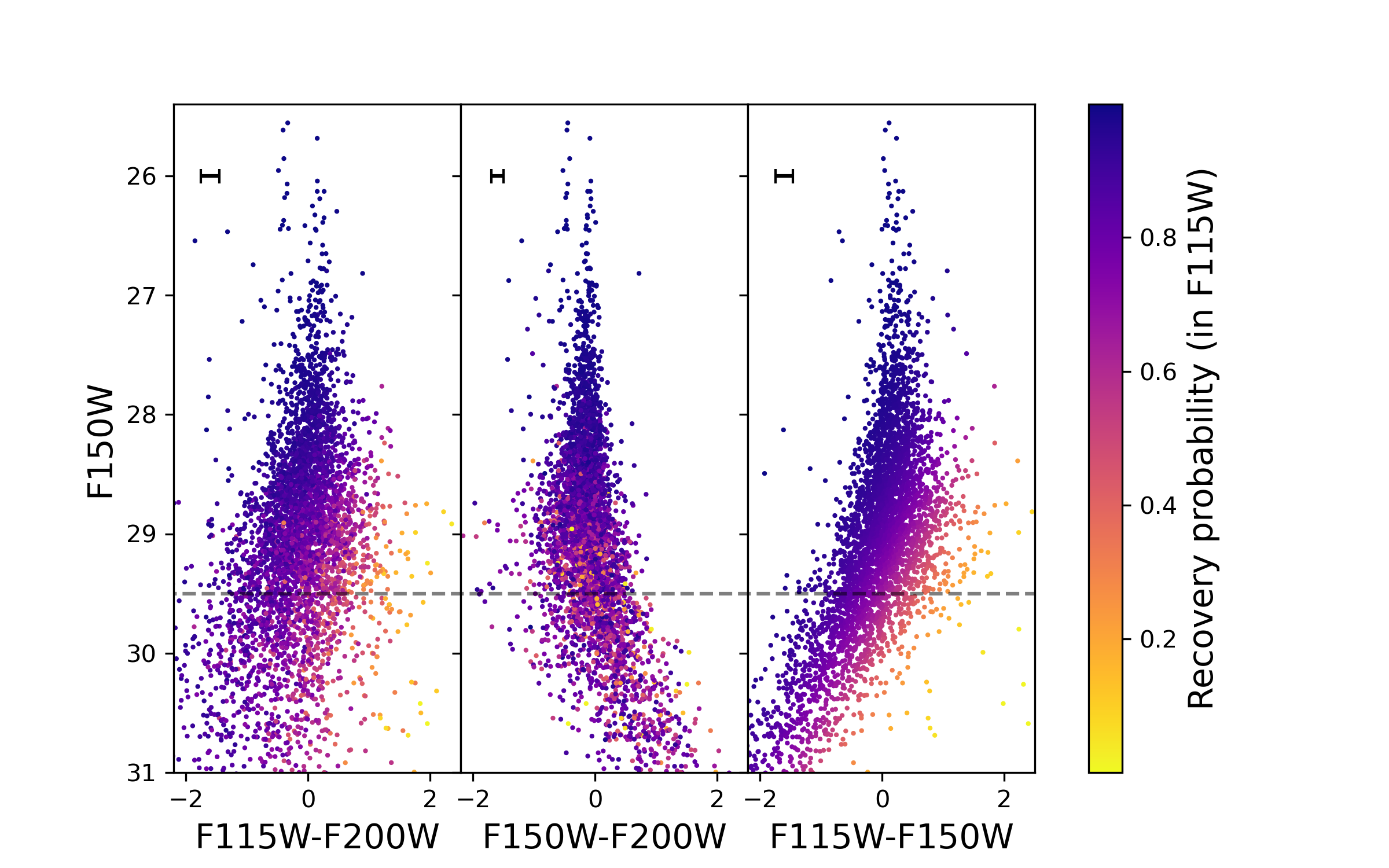}
\caption{Color-magnitude diagram of point sources in
RXJ 2129. Magnitudes are in AB magnitudes corrected
for the cosmological K correction and foreground
reddening. Points are color-coded by the recovery
probability using the logistic regression fit. The dashed line is the 50\% recovery level.
\label{fig:CMD_recov}}
\end{figure}

We present color-magnitude diagrams of all candidate GCs in Fig \ref{fig:CMD_recov}, color-coded according to the recovery probability in F115W of each object.
Some of the faintest objects that passed through the culling steps may be compact background galaxies, and we also remove foreground stars as objects brighter than magnitude 25.5.
The remainder of point sources are the GCs plainly clustered around the BCG. We are left with 2720 GC candidates in the magnitude range $25.5 <$ F150W $< 29.5$. Weighted by the inverse recovery probability, and excluding objects outside of our completeness limit, our final count is 3162 GCs in our field of view of RXJ 2129. This is easily comparable to the 1000+ GCs observed in other BCGs \citep{Harris_2023BCGs}.

\section{Radial distribution} \label{sec:radial}

\begin{figure}[ht!]
\includegraphics[width=0.47\textwidth]{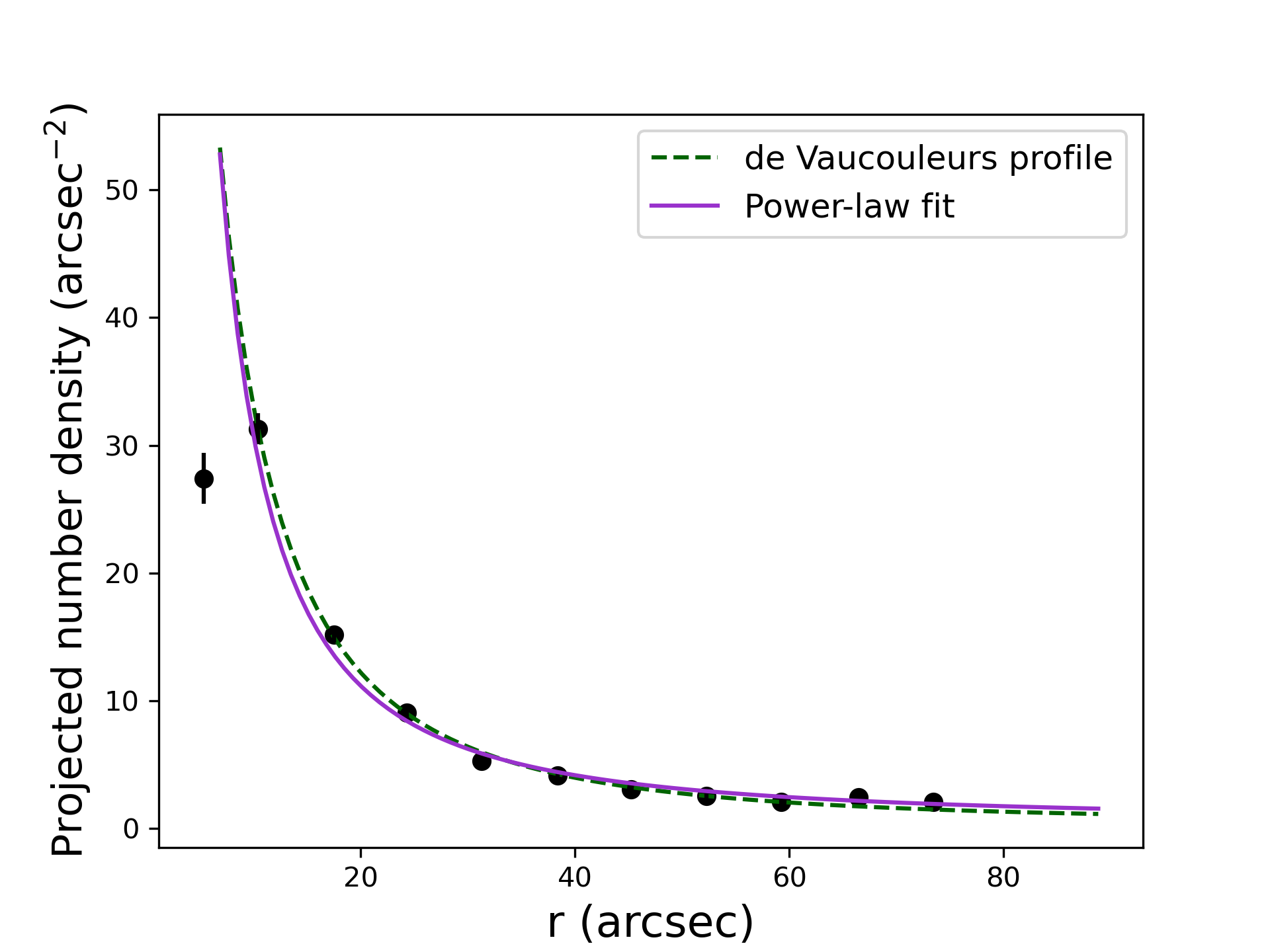}
\caption{Radial distribution of the GCs fit with a power-law and a de Vaucouleurs profile. Each GC count is weighted by it's inverse recovery probability, and objects with a probability of less that $50\%$ are removed. 
\label{fig:recov_radial}}
\end{figure}

The spatial distribution of GCs in RXJ 2129 is of particular interest. For major galaxies, GCs typically follow a spatial distribution well described by a power-law or Sersic profile.
We plot the projected number density as a function of distance from the center of the BCG with the least squares best fitting shown in Fig \ref{fig:recov_radial}.
When counting GCs, we weight each object by its inverse probability of recovery to calculate a projected object count in each radial bin. Any object with a probability of recovery less than $50\%$ in the F115W filter is removed. 
We exclude the innermost bin from the power-law fit as the BCG light limits our detection capabilities in this region. We find a power-law index of $1.58 \pm 0.04$. The background density is 0.5 arcsec$^{-2}$, consisting of a mix of distant GCs and faint contaminants that passed through the selection criteria.

For comparison, we also fit the distribution to a simple de Vaucouleurs profile. Shown in green in Fig \ref{fig:recov_radial}, this fit also visually matches the data quite well. We find an effective radius of $157 \pm 13$ kpc. 
\cite{Chu2022} investigated surface brightness Sersic profiles for a large collection of galaxies, including RXJ 2129. The authors found effective radii of 2.24, 44.48, and 190 kpc for the internal component of the BCG, the external component, and the intracluster light (ICL), respectively. Our GC distribution therefore has a larger effective radius than the stellar surface brightness, but a smaller effective radius than the ICL. 

With a virial radius of 1.3 Mpc \citep{Rines2013,Umetsu2018}, we can be confident that we are capturing the full radial extent of RXJ 2129's GC system. Fig \ref{fig:recov_radial} shows the fit of the radial distribution out to 290 kpc--easily enclosing the 10\% of the virial radius that is typically adopted for the spatial extent of GC systems \citep{Dornan2023,Dornan2025}. The effective radius of 157 kpc also lies within the expected range for BCGs \citep{Kartha2014,Hudson2018}.

\begin{figure}[ht!]
\includegraphics[width=0.5\textwidth]{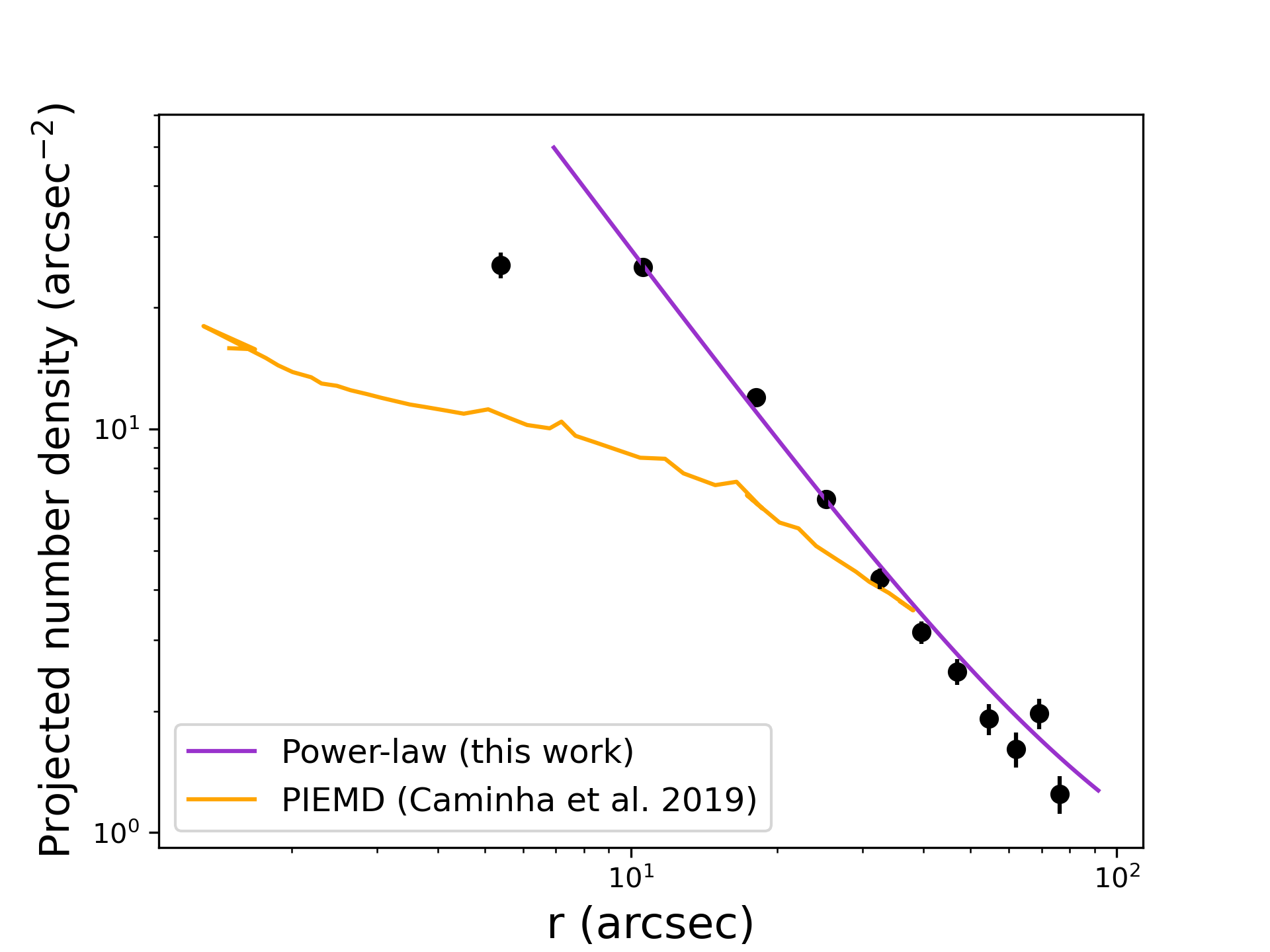}
\caption{Comparison of our power-law fit for the GC radial distribution to the mass profile modeled by \cite{Caminha2019}. The PIEMD profile was modeled in units of $M_\odot$ kpc$^{-2}$ and has no meaningful normalization to the GC projected number density. 
\label{fig:caminha}}
\end{figure}

Next, we compare our results to the total surface mass density profile modeled using strong lensing by \cite{Caminha2019}. 
Using multiply lensed background galaxies, the authors used a pseudo-isothermal elliptical mass distribution (PIEMD, \cite{PIEMD}) to model the smooth mass components. For RXJ 2129, all of the multiple images were within 100 kpc from the cluster center, which limits our comparison at larger galactocentric distances. We plot our comparison to their results in log-log space in Fig \ref{fig:caminha}, and conclude that the GC distribution is steeper than the total mass distribution. A potential explanation for this distinction is that the dark matter profile is shallower in the center of the cluster due to its collisionless nature, while the stellar component achieves greater central concentrations. 
However, the present data is limited to the BCG halo region and does not explore the GC distribution across the entire field of the cluster. 

\section{Color Populations} \label{sec:color}

In this section, we analyze trends and variations in relation to GC color. 
Through their chemical compositions, GCs record the conditions of the DM halos in which they formed.

GCs in large galaxies conventionally show a bimodal distribution in their color, which is closely related to their metallicity \citep{zinn1985,Barmby2000, peng2006,Strader2007,Harris2023}. 
The more metal-poor population exhibit a trend of bluer colors, older ages, and have a roughly spherical distribution. 
The more metal-rich GCs have redder colors, are younger, and are more centrally concentrated.
The history of the GCs is where this division originates: the more metal-rich population have been found to have mainly formed in more massive, enriched halos, while the metal-poor population formed ex situ and was later accreted \citep{Li2019, Keller2020, marta2022, chen2024}.  

As shown in Fig \ref{fig:CMD_recov}, there is no obvious division between the red and blue populations that would reflect a distinct metal-rich and metal-poor population in RXJ 2129.
While it is true that some systems have shown unimodal MDFs, another contributing factor is that near-infrared color indices are less sensitive to GC metallicity than are optical colors \citep{Harris2023}.
Additionally, there are significant measurement uncertainties in color that introduce scatter.  
For the present purposes, we simply divide the sample into red and blue subsets at a color of F115W$-$F200W$=0.018$, such that each has the same projected number of GCs. 


\begin{figure}[ht!]
\includegraphics[width=0.47\textwidth]{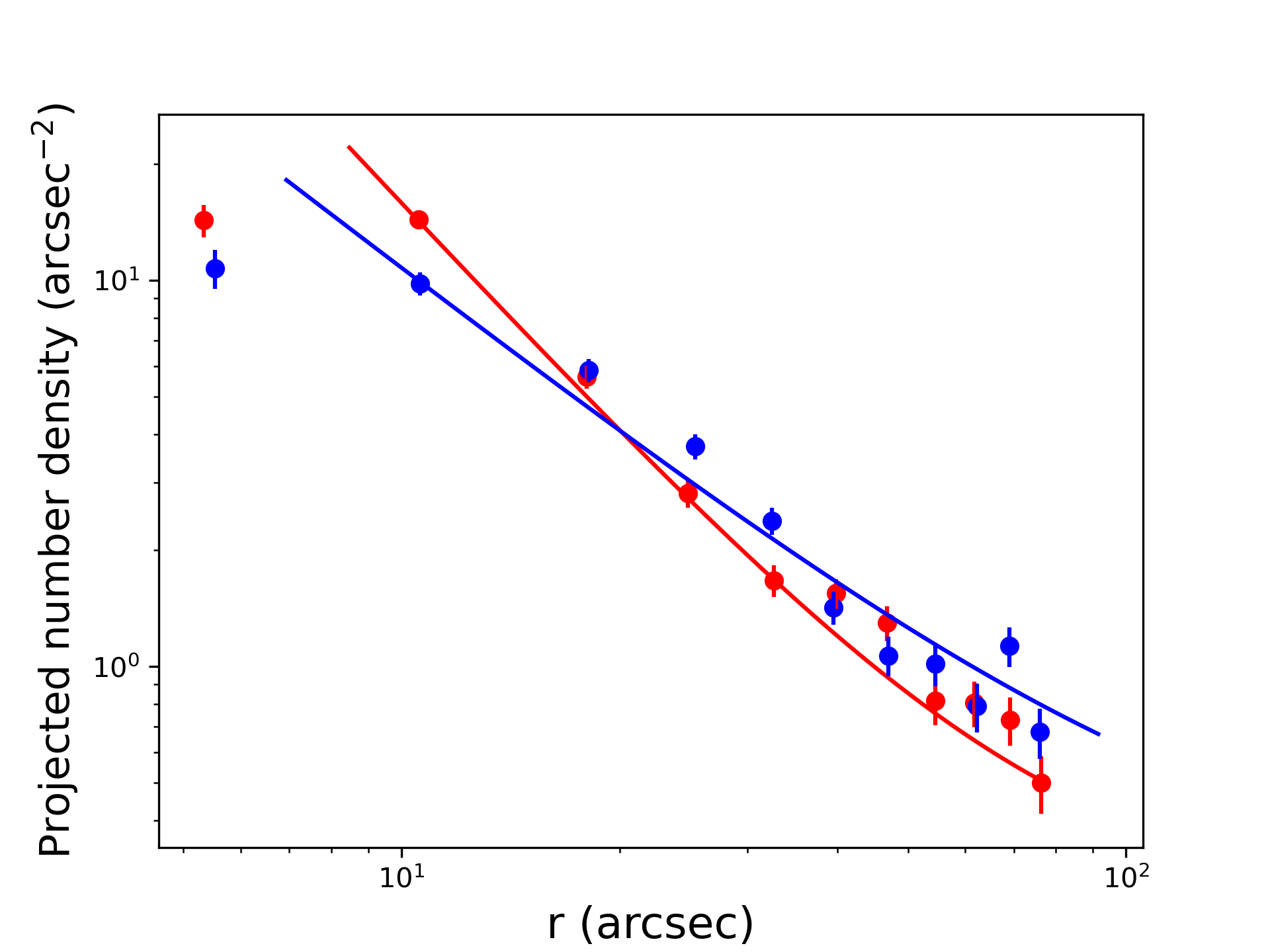}
\caption{Same as Fig \ref{fig:recov_radial}, but distributions are divided into red and blue sub-populations (with respective colors for data points and fit).
\label{fig:power2}}
\end{figure}

\begin{figure*}[ht!]
\centering
  \begin{subfigure}{0.45\textwidth}
    \includegraphics[width=\textwidth]{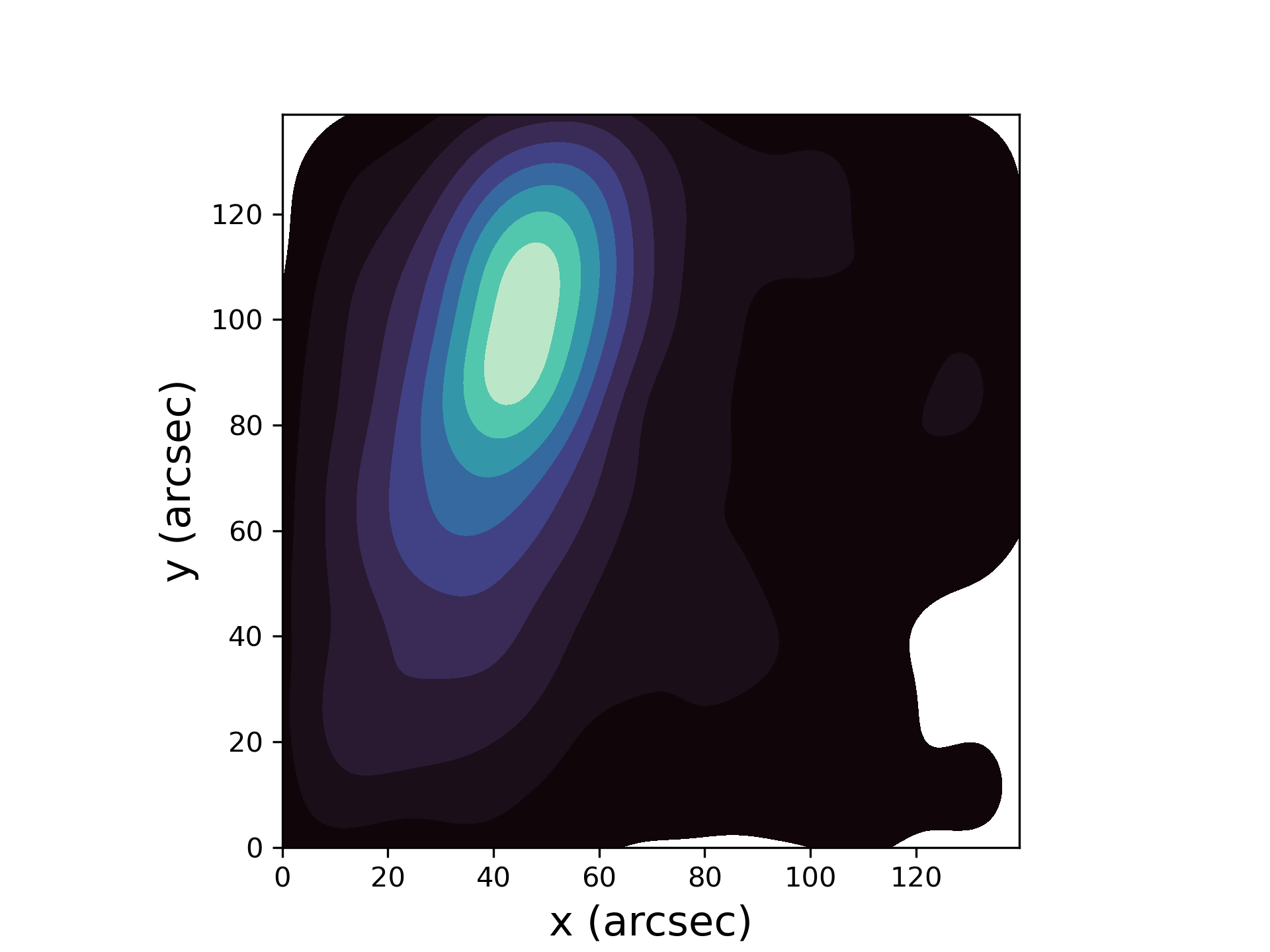}
  \end{subfigure}
  \hfill
  \begin{subfigure}{0.45\textwidth}
    \includegraphics[width=\textwidth]{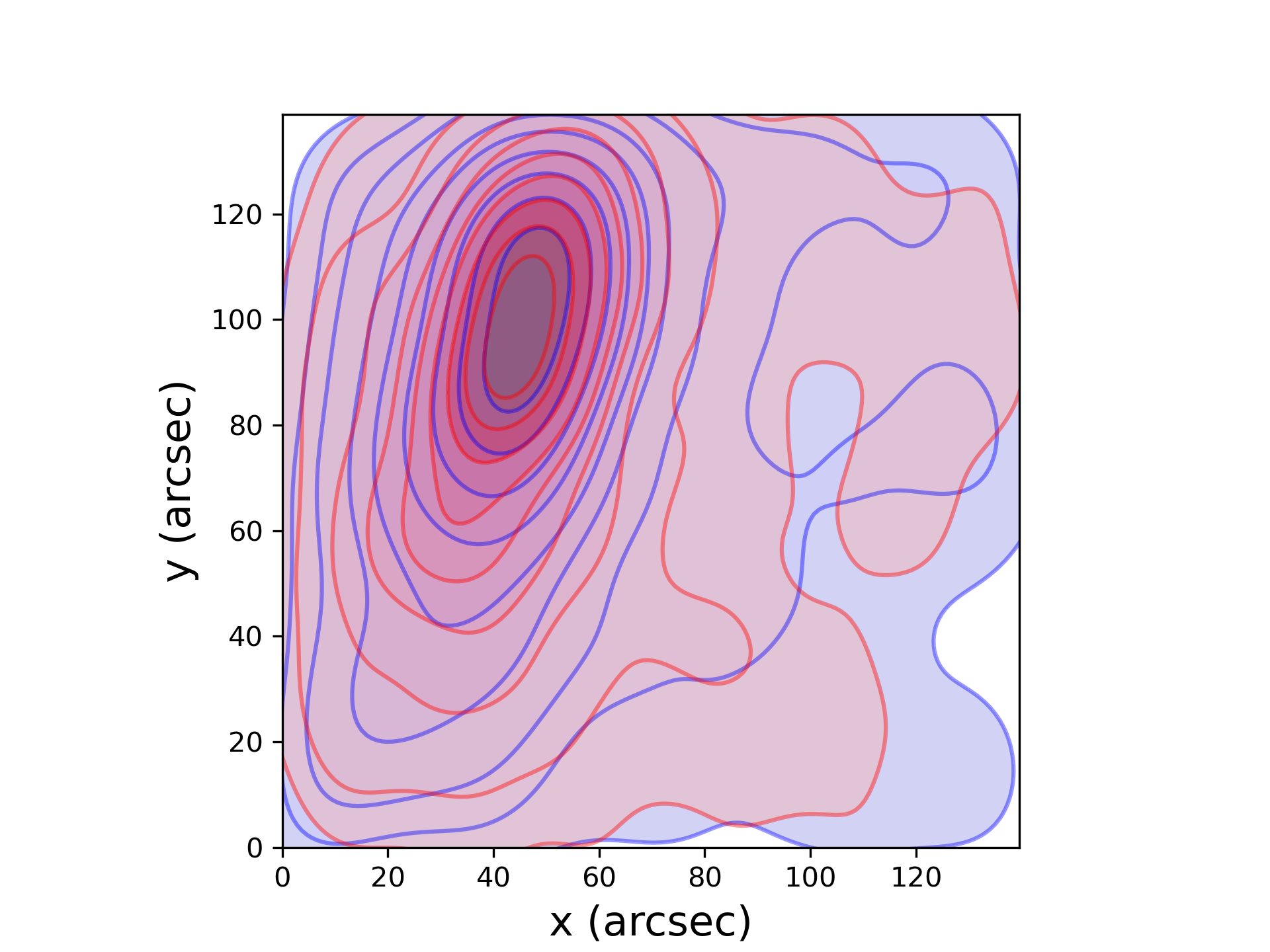}
  \end{subfigure}
\caption{Contours of globular cluster counts. The right plot shows the contours for the red and blue populations, and the left plot is the combined sample.   
\label{fig:contour}}
\end{figure*}

We repeat the radial distribution calculations and fitting on the red and blue sub-populations. The power-law indices recovered by the least squares fit are reported in Table \ref{table:ellip_param}. 
While a bimodality in color was not immediately visible, we do find significant distinctions when dividing the sample in half into a blue and red sub-sample and analyzing the spatial distribution. 
The redder population of GCs is more centrally concentrated, which agrees with the expectations as a population primarily formed in-situ. 
This is evident in the larger power-law index--reflected by the steeper slope in log space (Fig \ref{fig:power2})--and the right panel of the contour map in Fig \ref{fig:contour}.

\section{Ellipticity\label{subsec:ellipticity}}

As we can see in Fig \ref{fig:contour}, the distribution of globular clusters is highly elliptical, as is the BCG light profile itself. Our analysis in the previous section assumes circular bins, but we would now like to account for elliptical annuli. We can determine the ellipticity and position angle of the BCG using the expression for GC distribution given by \cite{ellip1994}:

\begin{multline}
    \sigma (R,\theta)=kR^{-\alpha}[\cos^2(\theta - \theta _p)+(1-\epsilon)^2 \sin^2(\theta - \theta_p)]^{-\alpha/2} \\ \equiv kR^{-\alpha}f(\theta)
\end{multline}

where $\theta_p$ is the position angle of the semi-major axis, $\epsilon$ is the ellipticity, $k$ is a normalization constant, and $\alpha$ is the power-law index obtained from the number density fit. To isolate the angular component,  we plot $f(\theta)$ as a function of position angle $\theta$:

\begin{equation}
    f(\theta)=[\cos^2(\theta - \theta _p)+(1-\epsilon)^2 \sin^2(\theta - \theta_p)]^{-\alpha/2}
\end{equation}

GCs are counted and normalized within azimuthal bins using an inner radius of 3.69 arcsec and an outer radius of 24.60 arcsec. These values are chosen such that there is no significant influence from the incompleteness at the center of the BCG, and we are capturing the elliptical behavior without looking too far out into the intergalactic medium.

\begin{figure}[ht!]
\includegraphics[width=0.47\textwidth]{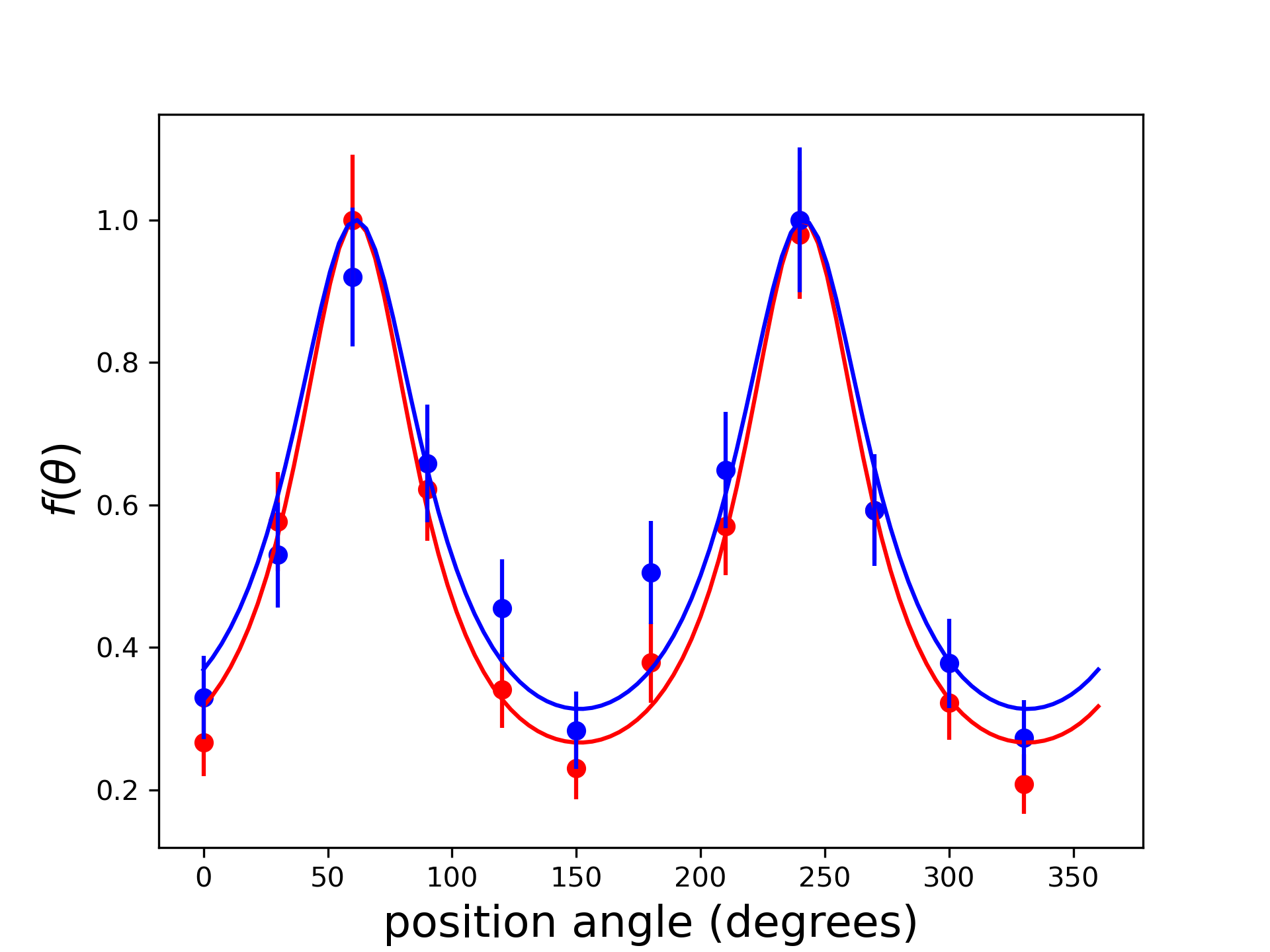}
\caption{Normalized azimuthal number density profile for red and blue GC samples. 
\label{fig:ellip}}
\end{figure}

We find an ellipticity of $0.54 \pm 0.01$ and a position angle of the semi-major axis of $61 \pm 2$ degrees North of West (counterclockwise from the positive x-axis) for the full sample. This agrees quite well with the ellipticity of 0.67 and position angle of 66.3 degrees for the stellar light determined by \cite{Caminha2019}. 
Once again, we repeat the process for the red and blue sub-populations; these values are reported in Table \ref{table:ellip_param}, and $f(\theta)$ is plotted in Fig \ref{fig:ellip}.

\begin{table}[h!]
\centering
\begin{tabular}{ |p{1.8cm}||p{1.5cm}|p{1.5cm}|p{1.8cm}|  }
 \hline
 Sample& Power-law index &Ellipticity &Position angle (degrees)\\
 \hline
 All sources   &$1.58 \pm 0.04$ &$0.54 \pm 0.01$ &$61 \pm 2$\\
 Red sources &$1.92 \pm 0.04$ &$0.56 \pm 0.01$ &$61 \pm 1$ \\
 Blue sources &$1.42 \pm 0.1$ &$0.51 \pm 0.02$ &$61 \pm 2$\\
 \hline
\end{tabular}
\caption{Parameters from the radial and angular best fits for the full sample of point sources, the red sub-sample, and the blue sub-sample.}
\label{table:ellip_param}
\end{table}

As expected, the red population has a greater ellipticity, reminiscent of the interpretation that that these GCs are primarily formed within the galaxy cluster and will more closely match the shape of the galaxies. 
The bluer population of GCs is less elliptical, which is consistent with the view that it has a higher proportion of accreted objects.
The position angle of the semi-major axis is consistent between the two populations.

\section{Comparison to Abell 2744}\label{sec:abell2744}

\begin{figure*}[ht!]
\centering
\includegraphics[width=0.9\textwidth]{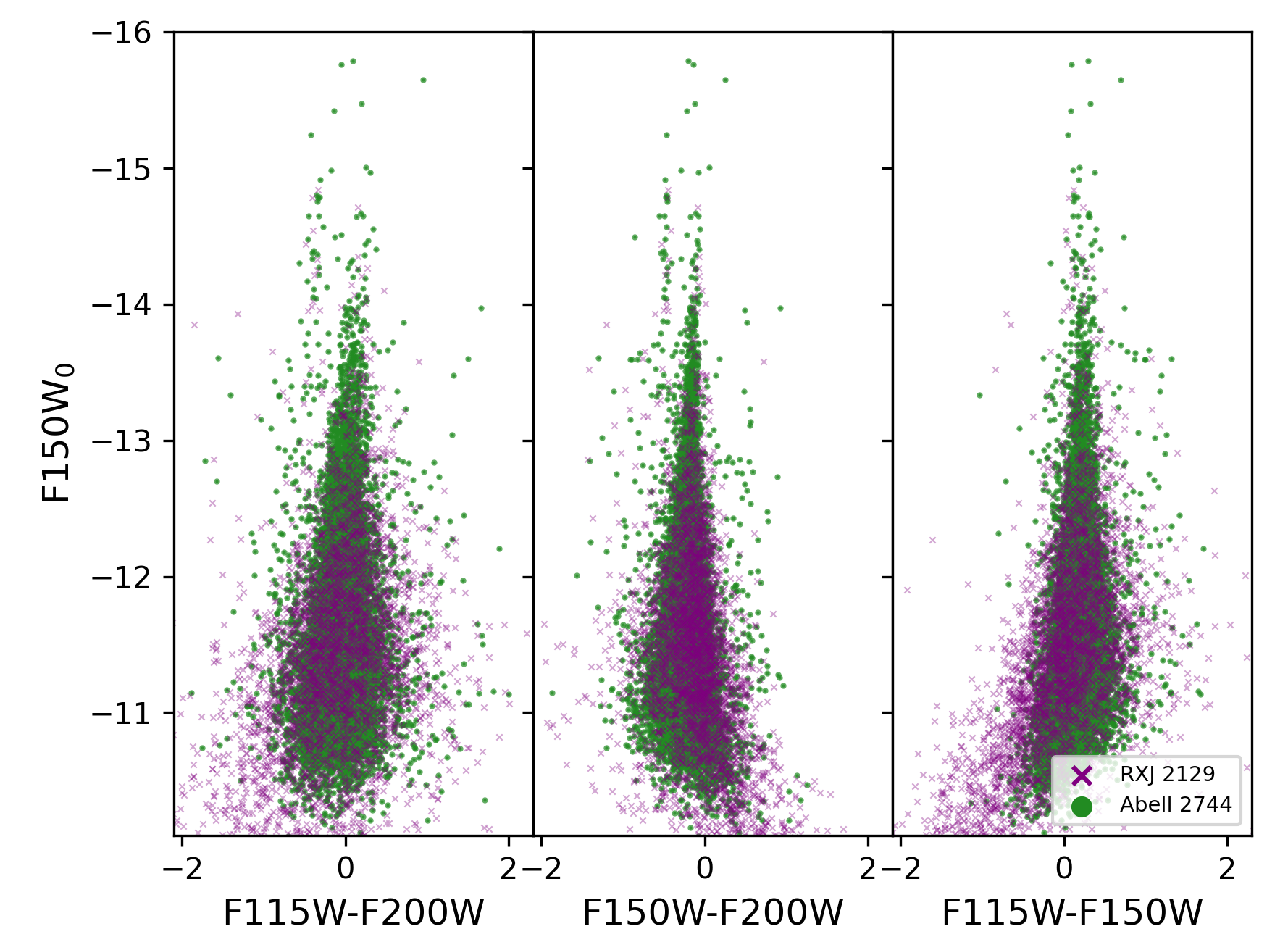}
\caption{Color-magnitude diagram of point sources in
RXJ 2129 and Abell 2744 \citep{Harris2023}.
\label{fig:abellRXJ}}
\end{figure*}

Abell 2744, another massive strong lensing galaxy cluster whose GCs have been intensively studied, has a redshift of 0.308, corresponding to a lookback time of 3.5 Gyr.
In Fig \ref{fig:abellRXJ}, we compare the absolute magnitude of the RXJ 2129 GC candidates to those in Abell 2744, determined by \cite{Harris2023}. 
The Abell 2744 population reaches higher-luminosity GCs, likely as a consequence of being a more populous system and having a younger stellar population by $\sim$ 0.6 Gyr.
On the other hand, dimmer luminosities are achieved in RXJ 2129.
The mean colors for the three indices are displayed in Table \ref{table:rxj_abell}; good agreement is seen between the two sets of data. 
RMS scatters around 0 in the three color indices average 118 for RXJ 2129 and 191 for Abell 2744. 

\begin{table*}[ht!]
\centering
\begin{tabular}{ |p{1.6cm}||p{2.5cm}|p{2.5cm}|p{2.5cm}|p{0.9cm}|}
 \hline
 System& (F115W-F200W) &(F150W-F200W) &(F115W-F150W) &N \\
 \hline
 RXJ 2129  &$0.05 \pm 0.01$ &$-0.20\pm 0.01 $ &$0.25 \pm 0.01$ &797 \\
 Abell 2744 &$-0.027 \pm 0.003 $ &$-0.219 \pm 0.004$ &$0.193 \pm 0.004$ &5281 \\
 \hline
\end{tabular}
\caption{Mean k-corrected color and error on the mean for RXJ 2129 and Abell 2744 \citep{Harris2023} objects brighter than F150W$_0$=-12.}
\label{table:rxj_abell}
\end{table*}

\begin{figure*}[ht!]
\centering
  \begin{subfigure}{0.48\textwidth}
    \includegraphics[width=\textwidth]{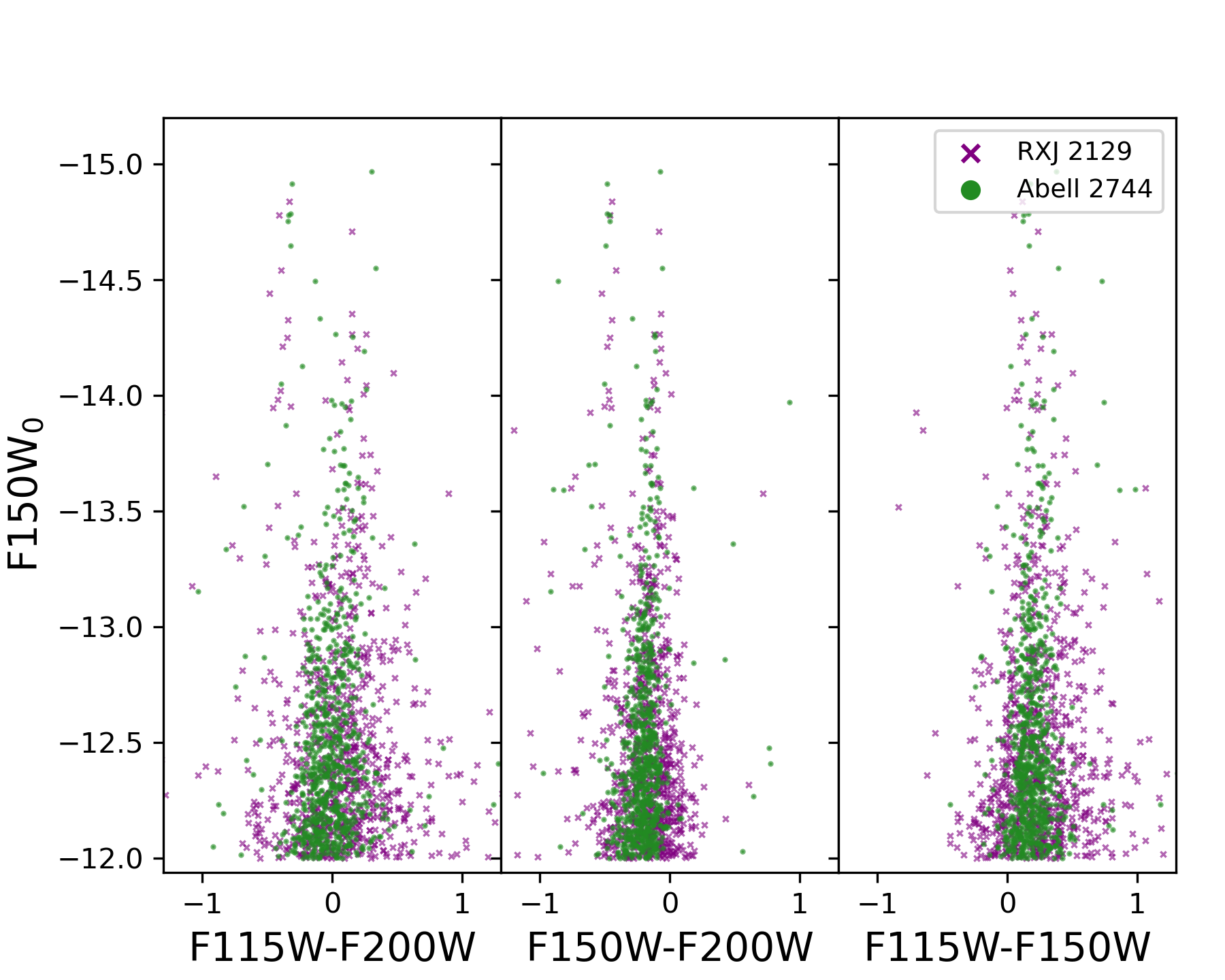}
  \end{subfigure}
  \hfill
  \begin{subfigure}{0.48\textwidth}
    \includegraphics[width=\textwidth]{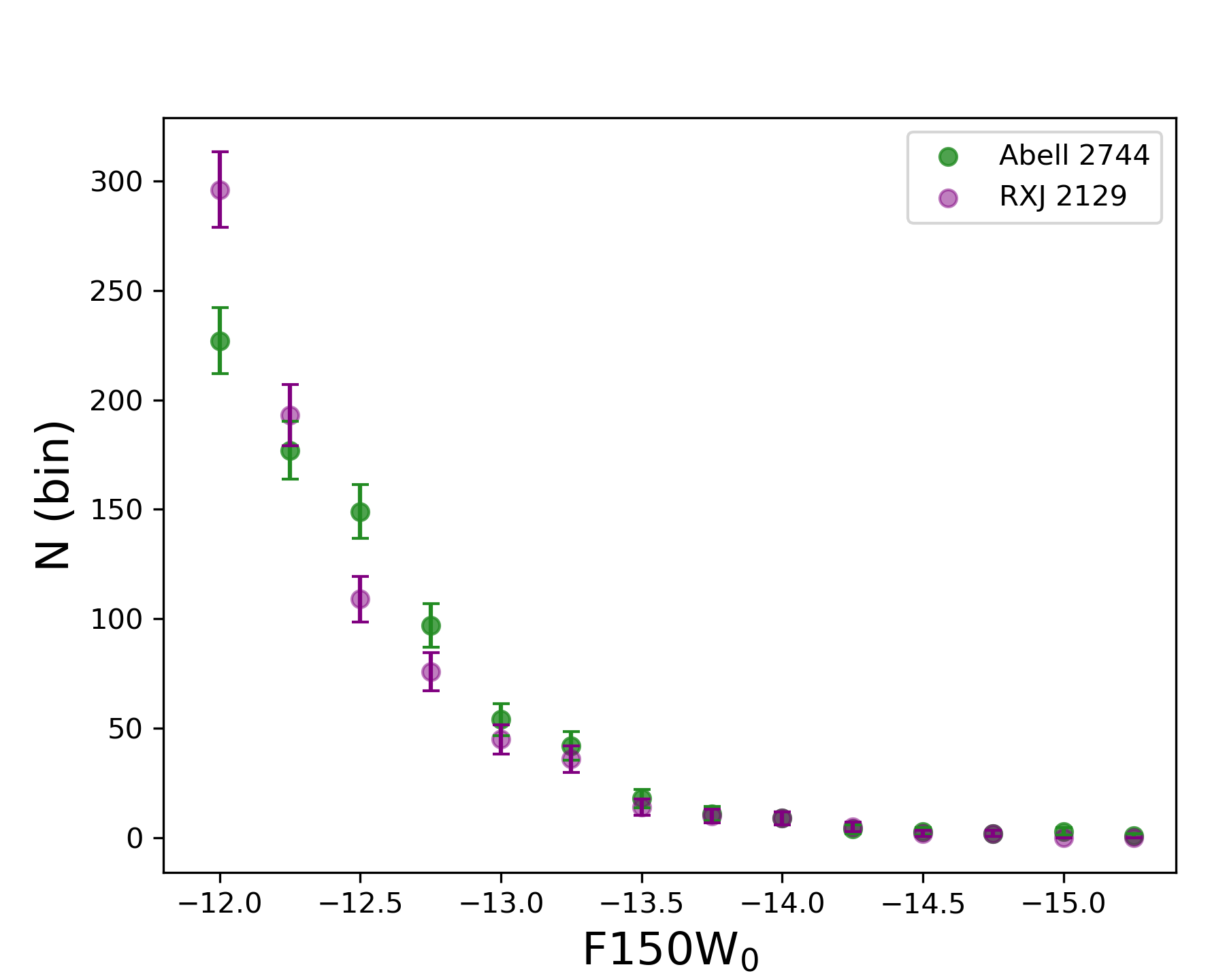}
  \end{subfigure}
\caption{Color-magnitude diagram of point sources in
RXJ 2129 and Abell 2744 for 797 objects (left) and number of objects observed in 0.25-mag bins in F150W$_0$ (right). 
\label{fig:luminosity}}
\end{figure*}

RXJ 2129 has 797 objects brighter than F150W$_0=-12$, only a fraction of the population observed in Abell 2744. To enable a fair comparison, we randomly select a subsample of 797 objects from Abell 2744. The color-magnitude diagrams and luminosity functions for both RXJ 2129 and this matched Abell 2744 sample are shown in Fig \ref{fig:luminosity}.
We note that our observations sample only the brightest end of the GC luminosity function (GCLF), well above the turnover point. The GCLF turnover is nearly universal at $10^5 L_\odot$ (corresponding to $M_{150}=-8.3$), with only a weak dependence on host galaxy mass \citep{Harris2014}. With a completeness limit of F150W=29.5, or $M_{150}=-10.8$, our luminosity limit is around $10^6 L_\odot$, meaning we are only detecting the most luminous GCs.

\begin{figure}[ht!]
\includegraphics[width=0.47\textwidth]{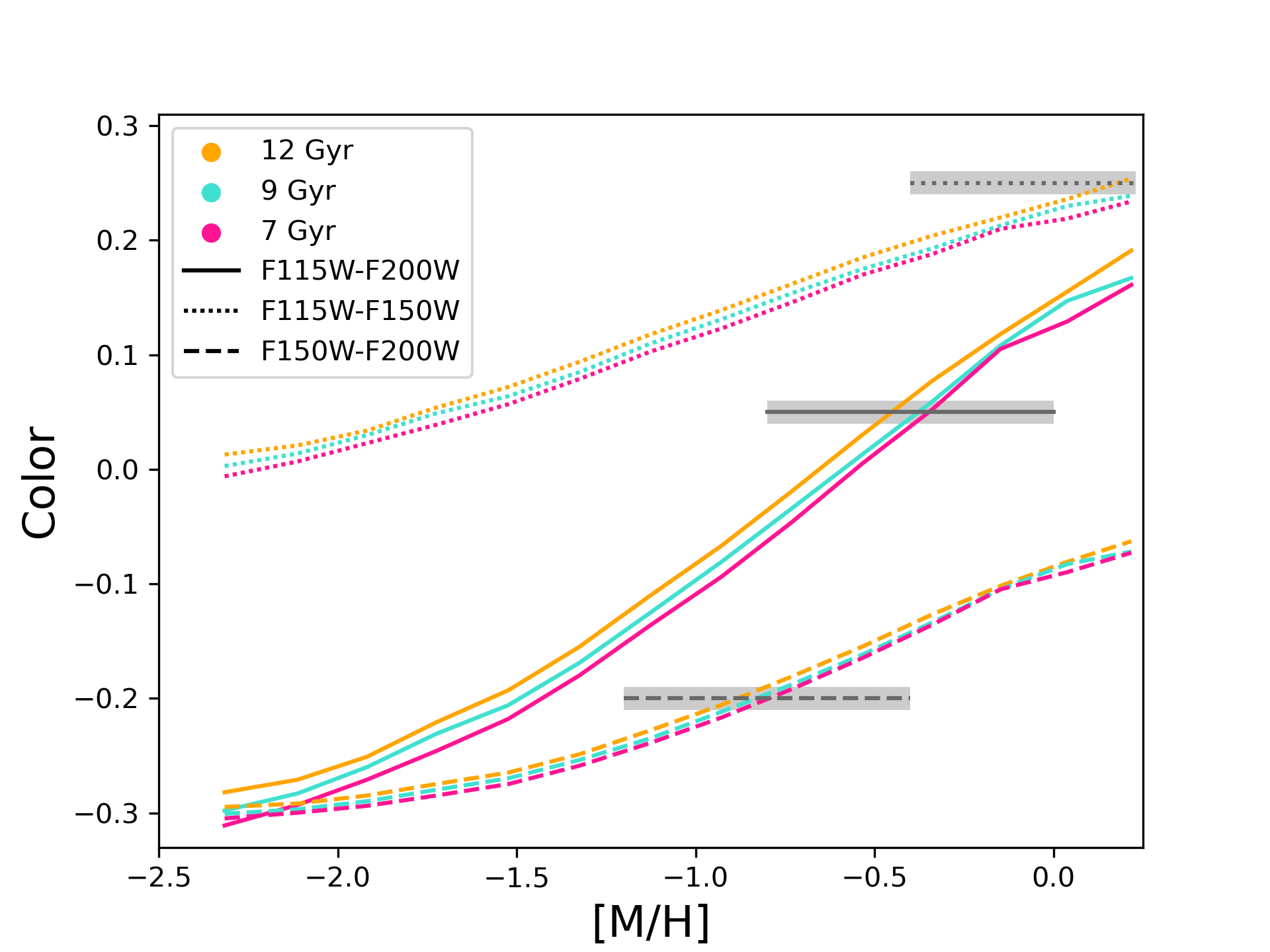}
\caption{Predicted color-metallicity relations from the PARSEC CMD3.7 single-burst stellar population models.
Colors and linestyles indicate the assumed ages and color indices (AB magnitudes). Grey lines indicate the mean color of RXJ 2129 GCs, as reported in Table \ref{table:rxj_abell}, with shaded regions indicating the errors on the mean.
\label{fig:isochrones}}
\end{figure}

\begin{figure}[ht!]
\includegraphics[width=0.47\textwidth]{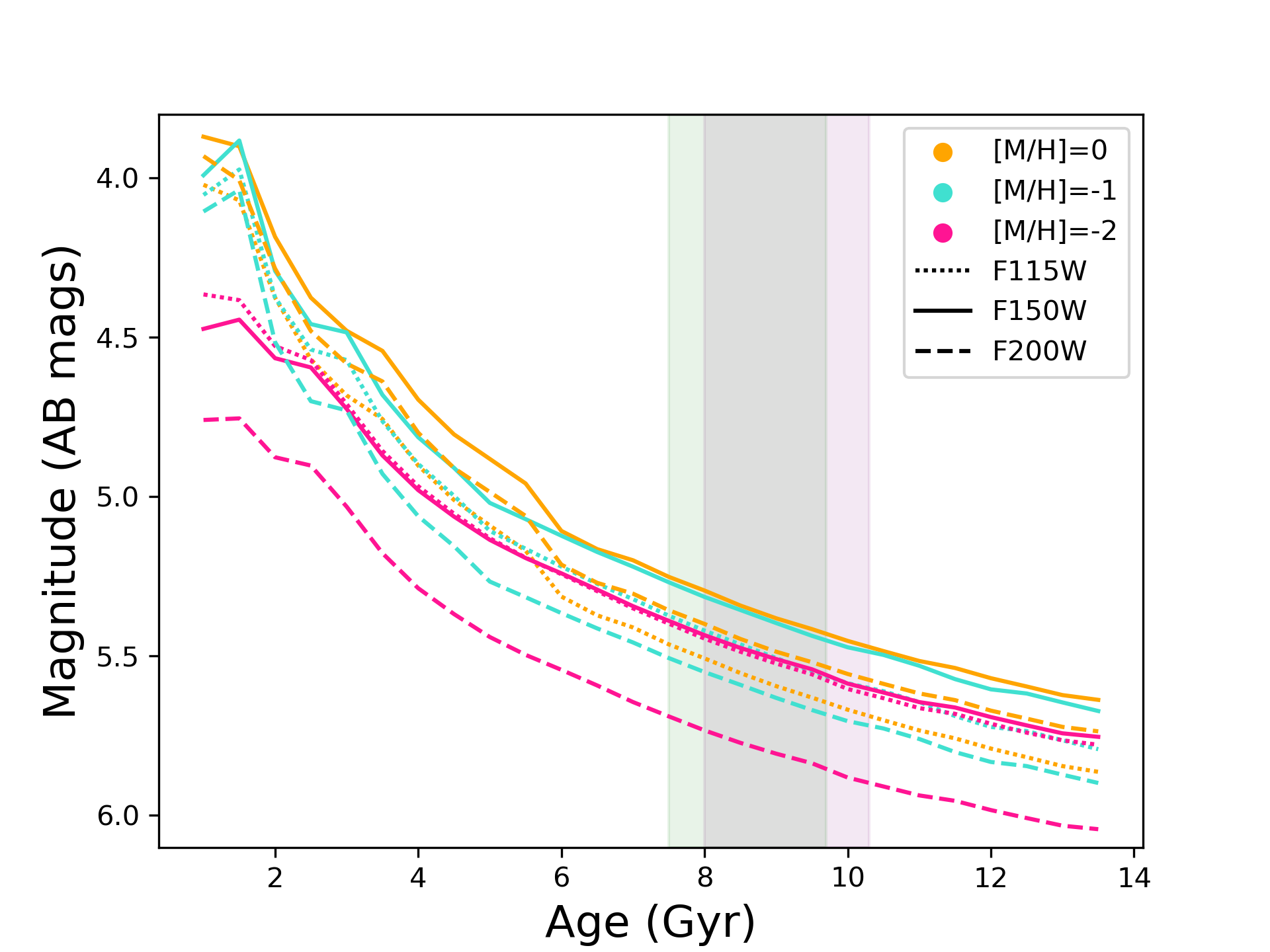}
\caption{Predicted luminosity-age relations from the PARSEC CMD3.7 models. Colors and linestyles indicate the assumed metallicity and NIRCam filter. Shaded regions show the expected age range of GCs in RXJ 2129 (purple) and Abell 2744 (green).
\label{fig:isochrone2}}
\end{figure}

In Fig \ref{fig:isochrones} and \ref{fig:isochrone2}, we show predicted color-metallicity and luminosity-age relations for GC populations of different ages and metallicities. These are generated using the PARSECv1.2S CMD3.7 single-burst stellar population models\footnote{\url{https://stev.oapd.inaf.it/cgi-bin/cmd_3.7}} \citep{parsec}. 
We can see that color (plotted in AB magnitudes) is fairly insensitive to age in these filters. 
Grey lines in Fig \ref{fig:isochrones} mark the intersection between the observed colors and the predicted color-metallicity functions; 
the mean metallicity of the RXJ 2129 GCs above F150W$_0=-12$ is estimated to be [M/H]$\sim -0.4$. We put the most reliance on the F115W-F200W filter, as it has the strongest relationship with metallicity. This value agrees with Abell 2744 within the combined uncertainties of the photometry and the K-corrections. 


Fig \ref{fig:isochrone2} shows that the relative evolution in luminosity with age is only weakly dependent on metallicity or filter choice, indicating that these bands are well-suited for comparing GC populations across different redshifts. 
Assuming an age range of 2-3 Gyr (as observed in Milky Way GCs, e.g. \cite{Leaman2013}), beginning $\sim$0.5 Gyr after the formation of the galaxy cluster, we highlight this interval as shaded regions in Fig \ref{fig:isochrone2} to represent the plausible age range of the GC populations.

The GCs of Abell 2744 and RXJ 2129 have very similar CMDs, luminosity functions, and metallicity within our completeness limit. 
The similarity despite their difference in lookback time is consistent with the idea that these populations formed at similarly early epochs, and have undergone little subsequent evolution besides a gradual decrease in luminosity over the additional 0.6 Gyr. A comparison over a larger range of ages would be possible if local GC systems were imaged in the JWST near-infrared wavelengths. 

\section{Summary and Conclusions} \label{sec:conclusion}

In this study, we have resolved and analyzed the population of GCs in the galaxy cluster RXJ 2129 at a redshift of 0.234. 
We conducted photometry on three images of RXJ 2129 from JWST NIRCam, and found 2720 GC candidates that appeared in the F115W, F150W, and F200W filters. 
We explored the completeness and photometric limits of our measurements by inserting artificial stars and fitting with a logistic regression function. 
Weighted by the inverse recovery probability of each object, our projected count is $3160 \pm 60$ GCs brighter than F150W = 29.5.

We have analyzed the radial and angular distribution of GCs from the center of the BCG, and compared to radial models of the surface brightness and total mass distribution predicted by previous studies on RXJ 2129. 
The distribution of the GCs around the BCG follows a radial power-law with an index of $1.58 \pm 0.04$. 
We also determined the ellipticity of the GC distributions and position angle of the semi-major axis to reflect the shape of the BCG.
When divided into a red and blue population, the redder GCs have a steeper power-law fit and a greater ellipticity. They are more centrally concentrated than their bluer counterparts, agreeing with predictions that redder GCs are primarily formed in situ while bluer GCs are primarily accreted. 

Our future goals are to compare the spatial distributions of the GCs in two dimensions to the total mass distribution. Using gravitational lensing maps, we can probe the gravitational field of the cluster, leading to further evaluation of
the strength of GCs as a tracer for dark matter in galaxy clusters. 

RXJ 2129 can be added to the growing series of giant galaxies in lensing clusters whose GCs can now be studied with deep JWST imaging. It will soon be possible to study such systems over lookback times extending up to 6 Gyr and beyond, and build up an observationally-based picture of GC system evolution.

\begin{acknowledgments}
    
We acknowledge financial support from the Natural
Sciences and Engineering Research Council of Canada
(NSERC). KK acknowledges support by the Ontario Student Assistance Program with an Ontario Graduate Scholarship. We would also like to thank Marta Reina-Campos for the helpful discussion and ongoing collaboration. 

\section*{Data Availability} 
 The JWST images of RXJ 2129 are publicly available at \doi{10.17909/q8pr-dt43}.

\end{acknowledgments}

\vspace{5mm}
\facilities{JWST(NIRCam)}

\software{Daophot \citep{daophot}, ds9 \citep{ds9}, IRAF \citep{IRAF1986,IRAF1993},  Matplotlib \citep{matplotlib}, NumPy \citep{numpy}, PARSECv1.2S \citep{parsec}, pyraf \citep{pyraf}, seaborn \citep{seaborn}, statsmodels \citep{statsmodels}
}

\bibliography{maindoc}{}
\bibliographystyle{aasjournalv7}

\end{document}